\documentclass[12pt,preprint]{aastex}

\begin{document}

\def\arcsecpoint{$''\!.$}
\def\arcminpoint{$'\!.$}
\def\deg{$^{\rm o}$}
\def\ltsim{\raisebox{-.5ex}{$\;\stackrel{<}{\sim}\;$}}
\def\gtsim{\raisebox{-.5ex}{$\;\stackrel{>}{\sim}\;$}}

%\journalid{}{}
%\articleid{}{}

%\slugcomment{submitted to {\it Publications of the Astronomical Society of the Pacific }}

\shortauthors{Dunn, et al.}
\shorttitle{Survey of Intrinsic Absorption with FUSE}

\title{A Survey of Intrinsic Absorption in Active Galaxies using the Far Ultraviolet Spectroscopic Explorer\altaffilmark{1}}

\author{Jay P. Dunn\altaffilmark{2},
D. Michael Crenshaw\altaffilmark{2},
S. B. Kraemer\altaffilmark{3},
\& J. R. Gabel\altaffilmark{4}}

\altaffiltext{1}{Based on observations made with the NASA-CNES-CSA Far Ultraviolet 
Spectroscopic Explorer. FUSE is operated for NASA by the Johns Hopkins University 
under NASA contract NAS5-32985.}

\altaffiltext{2}{Department of Physics and Astronomy, Georgia State University
, Atlanta, GA 30303. Email: dunn@chara.gsu.edu, crenshaw@chara.gsu.edu}

\altaffiltext{3}{Institute for Astrophysics and Computational Sciences, 
Department of Physics, The Catholic University of America, Washington, DC 
20064; kraemer@yancey.gsfc.nasa.gov.} 

\altaffiltext{4}{Physics Department, Creighton University, Omaha, NE 68138}

\begin{abstract}

We present a survey of 72 Seyfert galaxies and quasars observed by the 
{\it Far Ultraviolet Spectroscopic Explorer (FUSE)}. We have determined 
that 72 of 253 available active galactic nuclei (AGN) targets are viable targets for detection 
of intrinsic absorption lines. We examined these spectra for signs of intrinsic 
absorption in the O VI doublet ($\lambda$$\lambda$1031.9, 1037.6) 
and Lyman $\beta$ ($\lambda$1025.7). The fraction of Seyfert 1 galaxies 
and low-redshift quasars at $z \lesssim$ 0.15 that show evidence of intrinsic 
UV absorption is $\sim$50 \%, which is slightly lower than Crenshaw et 
al. (1999) found (60\%) based on a smaller sample of Seyfert 1 galaxies 
observed with the {\it Hubble Space Telescope (HST)}. With this new fraction 
we find a global covering factor of the absorbing gas with respect to the central nucleus 
of $\sim$0.4. Our survey is to date the largest searching for intrinsic UV absorption 
with high spectral resolution, and is the first step toward a more comprehensive 
study of intrinsic absorption in low-redshift AGN.

\end{abstract}


\section{Introduction}

Seyfert galaxies are relatively nearby, mostly spiral galaxies that host 
active galactic nuclei (AGN). Intensive studies over the past few decades 
lead us to believe that a supermassive black hole with an accretion disk 
lies at the core of every active galaxy and is the engine driving the 
activity. Seyfert galaxies, unlike quasars, are typically close, with 
redshifts $z \lesssim$ 0.1 and have moderate bolometric luminosities (10$^{45}$ ergs s $^{-1}$. 
Over time scales ranging from days to years Seyfert galaxies show variation in continuum 
luminosity over a factor of $\sim$10 in amplitude (Dunn et al. 2006).
Seyfert galaxies are divided into two basic categories, 1 and 2 (Kachikian 
\& Weedman 1974). Seyfert 1 galaxies show both broad permitted emission lines and 
narrow permitted and forbidden lines, while the UV and optical spectra of 
Seyfert 2 galaxies are devoid of broad emission features in unpolarized light. 

Oke \& Sargent in (1968) found that while the optical spectra of Seyferts 
have absorption features, which were previously attributed as stellar, there was an 
He I line in NGC 4151 that was likely due to self-absorption. Anderson \& Kraft (1969) 
found 3 kinematic components within the He I and hydrogen 
Balmer absorption lines that showed blueshifts relative to the rest frame
of up to 970 km s$^-$$^1$. They attributed these features to an outflow of 
gas from the core with multiple kinematic components. 

Cromwell \& Weymann (1970) showed that the absorption was 
variable. Ulrich \& Boisson (1983) found in data from the {\it International 
Ultraviolet Explorer (IUE)} that only 3 to 10\% of Seyfert galaxies showed
intrinsic absorption in high-ionization lines (C IV, N V, O VI). Crenshaw 
et al. (1999) found in an {\it HST} study 
that this number was far too low due to the low resolution of {\it IUE} and
that the percentage of Seyfert galaxies that showed intrinsic C IV absorption 
was closer to 60\%. This however was not a large survey, 17 objects, with 
possibly some selection biases. It did show that intrinsic absorption was 
much more prominent in Seyfert 1 galaxies than previously thought and that further study is 
required of this phenomenon to more fully understand the central engine in
AGN. Crenshaw et al. also showed a 1:1 relationship between the warm X-Ray
absorbers (George et al. 1998, Reynolds 1997) and intrinsic UV absorption in Seyfert 
galaxies. They also estimated a global covering factor for Seyfert galaxies
of $\sim$0.5. Laor \& Brandt (2002) surveyed 50 AGN that included both quasars 
and Seyfert galaxies. Their survey found that 44\% showed C IV absorption 
with an equivalent width greater than 0.1 \AA.

In a recent review article, Crenshaw, Kraemer, \& George (2003) found that while there 
have been many advances in the field of intrinsic absorption, the need for
further examination of mass outflow properties in AGN is necessary. We need
to have a larger survey of AGN to investigate the effects of luminosity, 
AGN type, radio power, orientation and accretion rate. Also, there is a 
need to determine transverse velocities of the intrinsic absorbing clouds
(as seen in Kraemer et al. 2001). These factors will help constrain dynamical 
models currently being considered as explanations of the origin of the 
mass outflow. This paper will be the first in a series designed to 
further our understanding of these characteristics. The main goal of the paper
is to presents the spectra, component identifications, and the frequency of 
occurrence of intrinsic absorption.

\section{Survey}

Our survey is more than 4 times larger than that of Crenshaw et al. 1999 and is 
taken from the available {\it FUSE} data at the {\it Multimission Archives 
at Space Telescope (MAST)} (url: http://archive.stsci.edu/). {\it FUSE} is ideal for intrinsic absorption 
studies in AGN due to the wavelength coverage (905 \AA\ to 1187 \AA) that
allows for the O VI doublet ($\lambda$$\lambda$1031.9, 1037.6) and 
Lyman $\beta$ ($\lambda$1025.7) to be detected at low redshifts. {\it 
FUSE} also lends itself to this work because of its resolution, 
approximately 15 km s$^-$$^1$ (FWHM), allowing us to find narrow absorption 
features and resolve structure in broader features. 

{\it FUSE} is comprised of 4 mirrors and 4 gratings split on to two 
detectors (Sahnow 2002) . This provides 8 different spectra per 
observation. One set uses a LiF coating while the other a SiC coating. The 
LiF coating provides a reflectivity nearly twice that of the SiC at 
wavelengths greater than 1050 \AA. In nearly every spectrum, this implies a
better signal-to-noise across the LiF spectrum in the region of interest. 
We downloaded the raw data for all of the targets we selected and processed 
them using CalFUSE v3.1 in time-tag mode (TTAG) (Dixon et al. 2002). We coadded 7 of the 8 
spectra into one spectrum weighted by exposure time. We did not include the 
LiF 1b segment in the coadded spectrum. We intentionally omitted this because
of a distortion (known as the "worm") in the spectrum for this segment that has no available correction.
For objects with multiple epochs of observing we averaged the spectra. In a future paper, we
will examine the absorption variability using the multi-epoch observations.

{\it FUSE} has no calibration lamp onboard, leaving wavelength calibration 
to ISM lines (Sahnow et al. 2000). This introduces a velocity error of 
approximately 20 km s$^-$$^1$ (Gillmon et al 2006), which is significant with 
a resolution of 15 km s$^-$$^1$. However our survey's initial purpose is to 
find Seyfert galaxies that exhibit intrinsic absorption and provide 
approximate velocity centroids. In a subsequent paper, we will present measured
velocities and velocity widths for all available absorption lines. Also, we will 
provide a good estimate of the
velocity error for any given spectrum based on the position of various ISM 
lines seen in spectra on a target-by-target basis.

Our list of targets originated from the category listing on the {\it MAST}
website. We took all targets listed by observers as Seyfert galaxies or as
Quasars, 253 targets. Using the NASA Extragalactic Database (NED) (url: http://nedwww.ipac.caltech.edu/index.html), 
we narrowed the total list by eliminating any targets that had a redshift
greater than $z \approx$ 0.15, to 143 objects. Any target with a redshift 
greater than this places the O VI doublet outside of the wavelength 
coverage for {\it FUSE}. We have retained these data for possible 
further analysis of the C III line ($\lambda$977.03) and the N III 
($\lambda$989.79) line.

We narrowed the list further to 122 objects, by removing any galaxy that 
had a type listed in NED other than Seyfert 1 or quasar; note that we still
examined the spectra of these targets, but these were not included in the 
survey. We include only Seyfert 1 galaxies or quasars because detection of 
intrinsic absorption requires a strong background source (i.e. continuum 
and broad line region). 

One interesting target that we removed is WPVS007. According to K. Leighly et
al. (2007, in prep), this Seyfert galaxy has evolved to a Broad Absorption 
Line mini-quasar. Thus, we have eliminated it from our list, although it 
has been known to show narrow intrinsic absorption lines in previous 
observations (Crenshaw et al. 1999).

The last criterion that we applied to the data was a signal to noise 
cutoff. By utilizing the provided noise vector with the data we found a 
good cutoff for the signal-to-noise per resolution element  ( $\sim$ 
15 km s$^{-1}$ FWHM) of 1.5 across the span 1050\AA\ to 1100\AA. Although
this value appears to be low, there are many resolution elements across a 
typical intrinsic absorption line. In this region we have data from the LiF 
portion of the spectrograph as well as data from the SiC portion coadded 
to give a good overall estimate of the quality of the spectrum. 

Table 1 lists our 72 AGN used in the survey. We list the signal-to-noise
we found in this region for a coadded spectrum per object in Table 1. We 
also provide a list of the observations available per object in Table 1. 
In the case of NGC 3516 we did not use all available observations. We chose
to use only the 2000 observation, ID P1110404, the LiF portion of the 
February 2006 observation and the January 2007 observation in our coadded 
spectrum. This set of spectra has a lower signal to noise, 1.9, than the 
total coadded spectrum, but these observations have no O I geocoronal 
dayglow interference, which is heavy in the O VI region of the spectrum.

\section{Absorption Detection}

\subsection{Identification}

In order to determine which lines are truly intrinsic versus lines that 
originate in our Galaxy's ISM, we used a program that provided a synthetic 
H$_2$ spectrum (Tumlinson et al. 2002), which we overlayed on our velocity 
plots. As an example, Figure 1 shows the spectrum of IRAS F04250-5718. We have the blue 
member and red member of the O VI doublet and the Lyman $\beta$ lines in 
respective order from top to bottom plotted in velocity space. The dashed 
line is the overlayed H$_2$ synthetic spectrum from the Tumlinson code, and
the other lines are ISM lines, taken from Morton (1991), scaled to their 
oscillator strength. The program requires H$_2$ columns to be input for 
each rovibrational level of the molecule. For detection purposes we input 
exceptionally high values for the column for each rovibrational level (lower 
angular momentum level J=0 through 7), on the order of 10$^1$$^9$ cm$^-$$^2$. 
This was to remove any speculation as to whether a line was truly H$_2$ in 
origin or intrinsic to the AGN/host galaxy. For H$_2$ lines that were 
coincident with an possible intrinsic absorption feature, we examined
nearby H$_2$ lines for both number and strength to place estimates on the
contamination.

Another factor that presented problems was geocoronal dayglow emission 
lines. These are background lines from the Earth's atmosphere that appear 
as narrow emission lines, which can have peak flux levels several times that 
of the surrounding continuum spectrum (Feldman et al. 2001). These lines 
are strong when {\it FUSE} is observing during the orbit day at low-Earth 
approach. One example is seen in NGC 4151 as Lyman $\beta$ and O I dayglow
near 1025 \AA\ in Figure 2.

We visually examined the spectra for lines that did not match the H$_2$ synthetic 
spectra or the ISM lines. These were flagged for further examination. We identified 
clean absorption features that had components at approximately the 
same velocity in at least 2 of the available lines (the 2 members of the O VI 
doublet and Lyman $\beta$). Any absorption line that aligned in velocity space
for 2 of the 3 available lines with at least 2$\sigma$ in equivalent width was
identified as intrinsic absorption. We recorded their approximate velocities based on
plots similar to Figure 1, which we provide in Table 2. An example is IRAS F04250-5718. 
We can see in Figure 1 
for IRAS F04250-5718 that intrinsic absorption appears in all three at relatively the 
same velocity with one broad component ranging between -300 and -100 km s$^{-1}$ and an 
additional component at approximately -100 km s$^{-1}$.

A handful of targets are still enigmas. These few targets show unexplained 
lines, but either do not agree in velocity or are possibly contaminated by ISM. 
Any that only contained one line or highly questionable lines were 
discarded as non-absorbers. We provide a target by target explanation in 
the Appendix.

\subsection{Comparison}

There have been two previous surveys of intrinsic UV absorption in low-z 
AGN, Crenshaw et al. (1999) and the preliminary survey by Kriss (2002). 
Crenshaw et al. had 17 targets in their C IV survey of data from the {\it Faint 
Object Spectrograph (FOS)} and the {\it Goddard High Resolution 
Spectrograph (GHRS)}. In the paper they labeled each of the targets as 
absorbers or non-absorbers, and we agree on all but one target of the 
17, I Zw 1. Kriss had 16 targets which he identified with {\it FUSE} 
spectra as intrinsic absorbers. We disagree on three of the 
targets, I Zw 1, Mrk 304 and Mrk 478.

I Zw 1 is a weak C IV absorber in the spectrum taken by {\it FOS}. In the {\it 
FUSE} observation, I Zw 1 is in a low flux state. Thus, the O VI absorption 
lines may be present, but we label this target as a non-absorber because the 
potential lines are much like the surrounding noise. There is a line at 
approximately 1800 km$^-$$^1$, but we only see it clearly in the O VI blue 
member. The other two are contaminated by H$_2$. Mrk 304 and Mrk 478 have 
both been categorized as absorbers by Kriss (2002), but in our observations
we find that Mrk 304 is only a possible absorber, and does not fit our conservative 
criteria. Mrk 478 shows no lines that cannot be explained by ISM contamination. 
Mrk 304 shows a possible line at 1800 km s$^-$$^1$ but it can only be seen in the 
O VI red member and Ly $\beta$. Thus by our standards, this object is only a possible 
absorber, and we do not include it as a firm detection. While NGC 7469 is a well known 
intrinsic absorber, the signal-to-noise of NGC 7469 combined with heavy H$_2$ 
contamination lead to a problematic spectrum.

Another survey we tested against was Laor \& Brandt (2002), which used 
mostly {\it HST} data. While they had a broad range of redshifts and object
types, there were a 14 objects that appeared in both of the studies. Of these 12 
objects we agreed on 9. One object appears in both studies, but the 
signal-to-noise in the {\it FUSE} spectrum is too low for our study. Once 
again I Zw 1 and Mrk 304 are not absorbers on our list, but Laor 
\& Brandt  has classified them as absorbers. Mrk290 however is listed by 
Laor et al. as a non-absorber, while we see a distinct absorbing feature at
approximately 200 km $^{-1}$.

\section{Conclusions}

For our final count we find of the 72 AGN with reasonably good signal-to-noise 
that 35 are intrinsically absorbing, with 11 new detections. Crenshaw et al. (1999) 
found approximately 60\% of their sample show intrinsic absorption. In our survey, 
we find that 49\% are absorbers. This could very well be a lower bound on the value 
due to the number of observations where the exposure times were not long enough to 
obtain a good signal-to-noise for the object. So the sample may be slightly
biased toward detecting absorption in AGN that are brighter in the far-UV. 

An important quantity to be determined is the global covering factor( C$_g$), 
which is the fraction of the sky covered by the ensemble of absorbers as seen 
from the nucleus. This quantity is useful in helping us understand the 
geometry of the nucleus and the absorbing region. C$_g$ = F$<$C$_f$$>$, 
where F is the fraction of galaxies that show intrinsic absorption and 
$<$C$_f$$>$ is the covering factor in the line of sight for the deepest component,
averaged over all AGN with absorption. 
To estimate C$_f$ for each individual AGN, we measured the residual
flux (F$_r$)
in the core of the deepest line. Assuming the component is
saturated, and the
residual flux is therefore unabsorbed continuum plus BLR flux
(F$_c$), then C$_f$
= (1 - F$_r$/F$_c$). If the core is not saturated, which is unlikely
in most
cases, the value of C$_f$ that we derived is actually a lower limit.
Crenshaw et al. 1999 found $<$C$_f$$>$ $\approx$ 0.85. We measured
the residual flux in saturated lines in the {\it FUSE} data and found that 
$<$C$_f$$>$ $\approx$ 0.86, very similar in both UV and far UV.
So using $<$C$_f$$>$ $\approx$ 0.86 with F $\approx$ 0.49 we find a global 
covering factor of C$_g$ = 0.42. 

This measurement has also been made for other AGN samples. Ganguly et 
al. (2001) found a fraction of only 0.25 for C IV absorption in quasars
with a z $<$ 1.0, while the Laor \& Brandt study found F = 0.50 for 
Seyferts and quasars up to redshifts of 0.5. George et al. (2000) found 
F = 0.3 for X-ray absorption in low-z quasars, and Vestergaard (2003) 
found F = 0.55 for quasars between 1.5 and 3.5 in redshift for C IV 
absorption. So for several luminosities and a wide range of redshift, most 
surveys find values for F around 1/2.

In our next paper, we will provide measurements by fitting these lines to 
find velocity centroids and equivalent width measurements. Once we have 
these values, we will examine each individual spectrum for variability in 
equivalent width, velocity and/or new components, which could lead to 
transverse velocities for the absorbers as seen in Kraemer et al. (2001).

\acknowledgments

This research has made use of the NASA/IPAC Extragalactic Database (NED)
which is operated by the Jet Propulsion Laboratory, California Institute of
Technology, under contract with the National Aeronautics and Space
Administration. This research has also made use of NASA's Astrophysics Data
System Abstract Service. We acknowledge support of this research under NASA
grants NNG05GC55G, NNG06G185G and NAG5-13109.

We also would like to thank Dr. Charles Danforth for his technilogical 
expertise, advice and progamming.

\appendix
\addcontentsline{tar}{section}{appendices}

\section{Notes on Individual Objects with Intrinsic Absorption}

\subsection{QSO 0045+3926}

QSO 0045+3926 is a new intrinsic absorption discovery. We find that this 
object shows clear single line absorption in all three lines. The O VI
region is found in the SiC portion of the spectrum; while the signal 
is lessened, there are no ISM lines nor any H$_2$ contamination.

\subsection{Ton S180}

Ton S180 shows very weak absorption, with some contamination from H$_2$ and
a N II ISM line in the Lyman $\beta$ line and no sign of H$_2$ absorption 
in the O VI red member. The Fe II ISM appears to be negligible based on 
the essentially non-visible nearby lines. This object was first seen as an 
absorber in Kriss (2002) but no intrinsic absorption was detected longward 
of 1200 \AA.

\subsection{Mrk 1044}

Fields et al. (2005) have identified the same lines we found in the {\it 
FUSE} observation. The obvious component appears in both the O VI 
lines, but the Lyman $\beta$ line is seriously contaminated via H$_2$. 
There is also evidence for a weaker line that appears to be slightly 
blended with Ar I in the O VI blue member, but is not clearly visible in 
O VI red and non-existent in Lyman $\beta$.

\subsection{NGC 985}

Absorption for this target was seen by Kriss (2002). Kriss lists this 
target with only one component, while we find that there are three narrow 
line absorbers with one blended broad absorber, which may consist of  
be two sub-components. The H$_2$ spectrum is varied in each of the O VI and
Lyman $\beta$ lines allowing for clear identification of each of these 
lines.

\subsection{EUVE J0349-537}

EUVE J0349-537 was first found in the EUVE survey and the optical 
counterpart was found by Craig and Fruschione (1997). Since that time it 
has appeared in 4 more surveys, but no exorbitant amount of study has been 
placed upon it. This target shows one absorber that is broad enough that it
is clearly above the noise, with a second nearby absorption feature that 
could be noise as it is only slightly visible in the O VI lines. This is a 
new absorption detection.

\subsection{IRASF 04250-5718}

This object originated in the Einstein Slew Survey (Elvis et al. 1992). 
We find at least two components absorbing in the far ultraviolet. It shows a 
broad component blended with a narrower component at a slightly lower velocity 
with very little contamination. This absorption has been seen previously by 
Kraemer et al. (1999).

\subsection{Mrk 79}

We see one component that is broad and shows very little contamination in 
the O VI lines. The Lyman $\beta$ line is significantly weaker but still visible
with little contamination. The second component is a narrow feature that 
is visible in the O VI red member, while the blue is highly 
contaminated and the L$\beta$ absorption is weak at best. This absorption has 
been identified previously. (Kraemer et al. 1999)

\subsection{Mrk 10}

This object shows H$_2$ interference in both of the O VI doublet lines,
and Ar I ($\lambda$1067) in the O VI red member. The absorption
features are much too broad for H$_2$ and ISM absorption to completely
explain them. Thus we find one or possibly two broad blending components.
Mrk 10 has not been identified before as an intrinsically absorbing object.

\subsection{IR 07546+3928}

This object was found in the New Bologna Sky Survey (Ficarra et al. 1985); 
it has been flagged for possible C IV absorption studies with the {\it 
International Ultraviolet Explorer (IUE)}  (Lanzetta et al. 1993). It has 
however not been labeled as an intrinsic absorbing target until now. In the
{\it FUSE} observations there are two broad components which are heavily 
contaminated in the O VI blue member by Fe II ISM lines. The Ly $\beta$ and 
the O VI red member are fairly devoid of contamination.

\subsection{PG 0804+761}

While the spectrum of this object shows dayglow N I, the intrinsic 
absorption lines are evident with no clear interference from ISM. This object
is a newly found intrinsic absorber.

\subsection{Ton 951}

Ton 951 was identified by Kriss in his 2002 conference paper. We find a single
narrow absorption feature. Only one of the three available lines has any 
H$_2$ contamination.

\subsection{IRAS 09149-62}

One of the less well-studied AGN targets, IRAS 09149-62 shows broad absorption
features across both members of the O VI doublet. The N II dayglow lines
leave whether or not Ly$\beta$ shows absorption to speculation. This detection
is a first for this object.

\subsection{Mrk 141}

This spectrum is highly noisy and the continuum flux level is extremely 
low. There are two lines that are in the same place in velocity space from 
O VI, but this object is not an ideal example of an intrinsic absorber due to the
lack of a Ly$\beta$ line at that velocity. It was classified as an 
intrinsic absorber by Kriss (2002).

\subsection{NGC 3516}

We present in this paper 2 new observations of NGC 3516 taken by {\it 
FUSE}. There have been a total of 6 observations taken with {\it FUSE}. In 
the first observation Kriss (2002) identified the same components seen in 
FOS, STIS and GHRS data by Crenshaw et al. (1999) and Kraemer et al. 
(2001). Kraemer et al. showed that NGC 3516 shows variability in the 
absorption features, allowing them to find a lower limit on the transverse 
velocity of $\sim$1800 km s$^{-1}$.

\subsection{ESO 265-G23}

This spectrum is full of H$_2$ features; however there is at least one 
absorption feature and two more possible components. The Ly$\beta$ for the
second possible component has a strong N II ISM line in it while the O VI 
red member is aligned with a Fe II ISM line. This is a new intrinsic UV 
absorption discovery.

\subsection{NGC 3783}

NGC 3783 was heavily studied by Gabel et al. (2005) using STIS and {\it 
FUSE} spectra. We find two broad components covering the span between -500 
and -800 km s$^{-1}$, which agrees with Gabel et al. (2005). However Gabel et al. 
found a component at -1350 km s$^{-1}$ which is coincident with a feature in 
Ly$\beta$; the O VI lines are somewhat less convincing in the {\it FUSE} 
spectrum.

\subsection{NGC 4051}

NGC 4051 shows a broad absorbing region that has no real contamination in 
the O VI blue member. The O VI red member is visible, but an O I 
dayglow line lies inside the trough and the Ly $\beta$ has a similar 
problem with an O I dayglow line alongside the Ly $\beta$ dayglow line.
Using STIS, Collinge et al. (2001) found two absorbing systems, one at 
$\sim$-600 km s$^{-1}$ and one at $\sim$ -2400 km s$^{-1}$. Each of these broad
components break into up to 8 smaller components in STIS spectra in the 
C IV and N V lines. In the {\it FUSE} data we see only the lower velocity 
component. There is no evidence in the {\it FUSE} spectrum for a higher 
velocity component.

\subsection{NGC 4151}

Kraemer et al. (2006, and references therein) performed an in depth study on 
NGC 4151, which showed multiple components in STIS spectra. In the {\it FUSE} 
data we find multiple components blended together due to the sensitivity O VI 
to intrinsic absorption. The Ly$\beta$ region is spoiled by the Ly $\beta$ 
dayglow. We only provide the centroid of the velocity in Table 2 for this object.

\subsection{RXJ 1230.8+0115}

Due to the high redshift of RXJ 1230.8+0115, there is little interference 
from the ISM (only 3 Fe II lines). There appears to be at least one broad
component with 2 other unexplained lines (1051 \AA\ and 1050 \AA) with 
two other broad absorption regions which do not
have corresponding matches in velocity space. This was recognized by 
Ganguly et al. (2001) and attributed to intervening gas in the IGM.

\subsection{TOL 1238-364}

This object has been classified as a Seyfert 2 galaxy (NED), but 
there does seem to be evidence for a broad line region in the {\it FUSE} 
spectrum. We find 1 broad component that shows some contamination from 
H$_2$, Ar I, and C I. Because this has been labeled as a Seyfert 2 galaxy,
it seems this object has not been considered as a target for intrinsic absorption
studies. 

\subsection{PG 1351+640}

PG 1351+640 has ample ISM contamination, but the overall absorption 
features look to be the same in all three lines. This was seen by Kriss 
(2002) and fitted by Zheng et al. (2001).

\subsection{Mrk 279}

Seen by Kriss (2002) and followed up with further study and Chandra 
observations by Scott et al. (2004), we find that there are three possible
components. Of the three only one is free from question of contamination.
The other two could be combinations of H$_2$ or ISM lines, however there
seems to be a paucity of H$_2$ absorption in nearby regions of the 
spectrum. Two of the three features appear to be intrinsic absorption.

\subsection{RXJ 135515+561244}

This spectrum is in a very low continuum flux state, but shows evidence
for one component of absorption with no contamination. This is a new 
intrinsic UV absorption discovery.

\subsection{PG 1404+226}

While the spectrum shows heavy ISM contamination, there are three 
absorption components uncorrupted in the O VI red member. These components 
agree in velocity with their O VI blue member and Ly $\beta$ counterparts 
and have very little overlap with the ISM lines. This was a common target
with the Laor \& Brandt (2002) survey.

\subsection{PG 1411+442}

Laor \& Brandt (2002) found that PG 1411+442 showed a large range of 
velocity for the intrinsic absorption ($\sim$5000 km/s). Unfortunately
with the redshift for this object, the O VI region falls into the SiC
portion of the spectrograph and provides less signal. We do not have 
enough signal to see the broad absorption Laor \& Brandt found,
but we have found two narrow components that are quite clear.

\subsection{NGC 5548}

NGC 5548 was found to be intrinsically absorbing by Shull \& Sachs (1993) 
in data from {\it IUE} along with evidence of X-Ray warm absorption (George 
et al. 1998). Crenshaw et al. (1999) found absorption in FOS data as 
well. More recently Crenshaw et al. (2003) saw five blended and broad 
components in data from STIS. We see a similar case as Crenshaw et al. did 
in the FUSE data with five blended and broad components.

Brotherton et al. (2002) examined the {\it FUSE} data available prior to 
2002 and found intrinsic absorption spanning the range between 0 and -1300 
km s$^{-1}$, with is coincident with what we have found in our coadded 
spectrum.

\subsection{Mrk 817}

We consider this a weak absorber. Mrk 817 shows a weak absorption line that is 
isolated in the O VI blue member, and has a weak Fe II contaminate in 
the O VI red member. The Ly $\beta$ is hardly visible, but this is 
easily understood due to the lower sensitivity of Ly $\beta$. This 
object has the fastest radial velocity component in a Seyfert galaxy to date
(Table 2), as seen in Kriss (2002).

\subsection{Mrk 290}

Seen by Kriss (2002), the absorber is a narrow ($\sim$200 km/s FWHM) 
absorption component. There is some possibility for contamination in each 
of the three lines, however the amount of contamination can be estimated 
by nearby ISM features. Because the nearby features have larger oscillator 
strengths and are weak lines, this contamination must be small in our 
absorption features.

\subsection{Mrk 876}

Due to its higher redshift (0.129), the O VI doublet falls in the SiC
region of the {\it FUSE} detector. But we still see a O VI broad line,
with a few narrow absorption features that are most likely intrinsic, 
and have not been seen before.

\subsection{Mrk 509}

Mrk 509 shows absorption broad enough that it was first seen in data from 
{\it IUE} by York et al. (1984). It also shows evidence for an X-Ray 
warm absorber, as discussed in Reynolds (1997) and George et al. (1998).
Kriss (2002) published the {\it FUSE} spectrum and identified the 
absorption components. 

Kraemer et al. (2003) examined STIS data and found 8 components spanning
the velocity range between -422 and +210 km s$^-$$^1$. They performed 
photoionization models of the absorbers and found that they are not the 
same absorbing regions as the X-Ray absorbers. 

In the {\it FUSE} observations with high resolution, in the Ly $\beta$ 
absorption feature we see the two same broad absorbers, but due to the
fact that the Ly $\beta$ line is less sensitive, we can see that the two 
broad components are between 4 and 5 components, the fifth overlapping 
with a coincident H$_2$ line.

\subsection{II Zw 136}

Crenshaw et al. (1999) found two components in the FOS spectra for 
II Zw 136, which we see repeated in our {\it FUSE} observations. Component
1 is clearly visible and virtually free of ISM interference; component 2 
however in the {\it FUSE} observations is clearly seen only in the O VI 
red member. Ly $\beta$ is weak at best and the O VI blue member is heavily
contaminated with an Fe II line and two H$_2$ lines in the vicinity.

\subsection{Akn 564}

Akn 564 shows 2 broad and blended absorption components. H$_2$ lines
heavily populate the area, but are far too narrow to account for the 
absorption. Crenshaw et al. (1999) saw absorption in the FOS data at the
same central velocities we find. Romano et al. (2002) published the {\it
FUSE} spectrum for this target and also identified the same components
we find. Crenshaw \& Kraemer (2001) found that this was 1 of 2 Seyfert
galaxies that showed traits they characterized as a dusty lukewarm
absorber.

\subsection{IRASF 22456-5125}

This object was seen first in the ROSAT wide field survey and later 
observed in the EUV. X-Ray studies have found this target to be 
highly variable (Grupe et al. 2001), but no ultraviolet intrinsic 
absorption has been previously identified. In our spectra we find 5 'finger-like'
narrow absorption components with little to no contamination from the
ISM.

\subsection{MR 2251-178}

Ganguly et al. (2001), using FOS and STIS observations, 
found variability in the absorption in both velocity and column density. 
They found that the velocity of the component was $\sim$ -1300 
km s$^{-1}$, in data from STIS and FOS. We find that this object shows clear 
absorption in both Ly$\beta$ and the O VI red member. While the blue 
member has both H$_2$ and ISM contamination, the velocity overlap is 
adequate enough to say that a significant portion of the absorption 
feature is intrinsic to the object. 

The question for MR 2251-178 is whether the absorption is best aligned at 
-2000 km s$^-1$ where the Ly$\beta$ absorption is weak and possible just 
ISM contamination, or at -300 km s$^-1$ where the O VI red member is weak. 
This could be a case of coincidental alignment where there are absorbers 
at both of these velocities; thus two of the features will be broader than 
the third component. Because X-Ray studies have shown this to be highly 
variable in X-Ray absorption (Halpern (1984)), and shown to be variable in 
the UV (Ganguly et al. 2001) it seems likely that the velocity on the 
component has shifted either accelerating to -2000 km s$^{-1}$ or 
decelerating to -300 km s$^{-1}$.

%\subsection{I Zw 1}

%This object is a source of conflict between this survey and previous surveys.
%Crenshaw et al. (1999) found absorption in STIS data while I Zw 1 was in a 
%higher continuum flux state. The {\it FUSE} observation was taken during a 
%low state for this object and thus much of the spectrum resembles noise.

%\subsection{NGC 4395}

%This spectrum has Lyman $\alpha$ dayglow, heavy H$_2$ absorption and two
%ISM absorption features along with a low flux state and a moderate signal-
%to-noise (4.39). Thus observing any intrinsic absorption is difficult, but
%we find one line that aligns in velocity space which could be real.

%In previous surveys this object has been neglected. Only in Iwasawa et al.
%(2000) is this target considered for absorption in X-Ray study. This also
%lends to the trend that X-Ray absorption is linked to the UV and optical
%intrinsic absorption seen by Crenshaw et al. (1999).

%\subsection{Mrk 304}

%Kriss (2002) found Mrk 304 to be a Seyfert with intrinsic absorption 
%in the {\it FUSE} data. As previously discussed this target has heavy 
%contamination from H$_2$ and we have placed it along with NGC 7469 as a 
%questionable target.

\subsection{NGC 7469}

NGC 7469 is a highly studied Seyfert 1 galaxy. Many surveys and studies 
to date have classified this as an intrinsically absorbing Seyfert galaxy (Scott
et al. (2005), Kriss (2002), Crenshaw et al. (1999)). Thus this object is 
a known intrinsic absorber. In the {\it FUSE} data however, the absorption 
appears very weakly and in the Lyman $\beta$ line there is a C II ISM line 
along with two heavy H$_2$ lines leaving doubt to the absorption feature. 
The O VI blue line is found at the same wavelength as Ar I ($\lambda$ 1048 
\AA) which tends to be a strong ISM line in the far UV.

\clearpage

\figcaption[fg1.eps]{Plot of the spectrum of IRAS 04250-5718 in velocity 
space. The intrinsic absorption is visible in components 
between $-$300 and $-$50 km s$^{-1}$. The ISM tick marks are sized relative 
to oscillator strength, and the H$_2$ spectrum is the dashed curve. Between 
$-$1000 and $-$500 km s$^{-1}$ there are possibly two ISM lines of Fe II that 
can be seen offset by approximately $-$50 km s$^{-1}$ with respect to the 
observed frame. This gives us a good approximation for all Fe II lines across 
the spectrum.}

\figcaption[fg2a.eps]{We present all of the spectra for the targets we 
identified as intrinsic absorbers. The spectra are plotted in the observed frame
and kinematic components of intrinsic absorption in the lines of Ly$\beta$ 
($\lambda$1025.7) and O VI ($\lambda$$\lambda$1031.9, 1037.6) are numbered. 
Broad components are indicated by brackets. Strong narrow emission lines are 
geocoronal. The dashed line represents the synthetic H$_2$ spectrum.
The specific velocity positions can be found in Table 2.}

\clearpage
\begin{figure}
\plotone{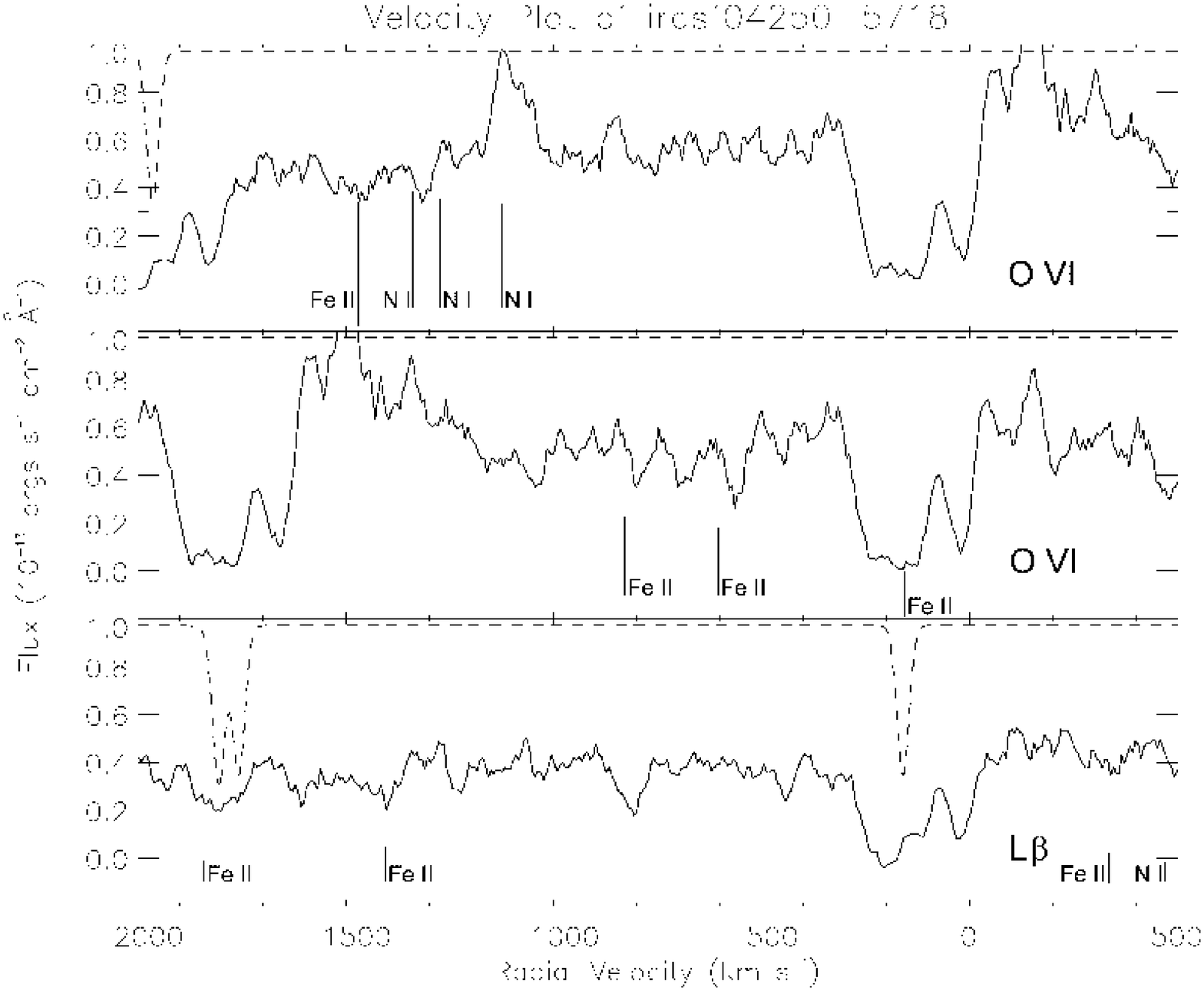}
\\Fig.~1.
\end{figure}

\clearpage
\begin{figure}
\plotone{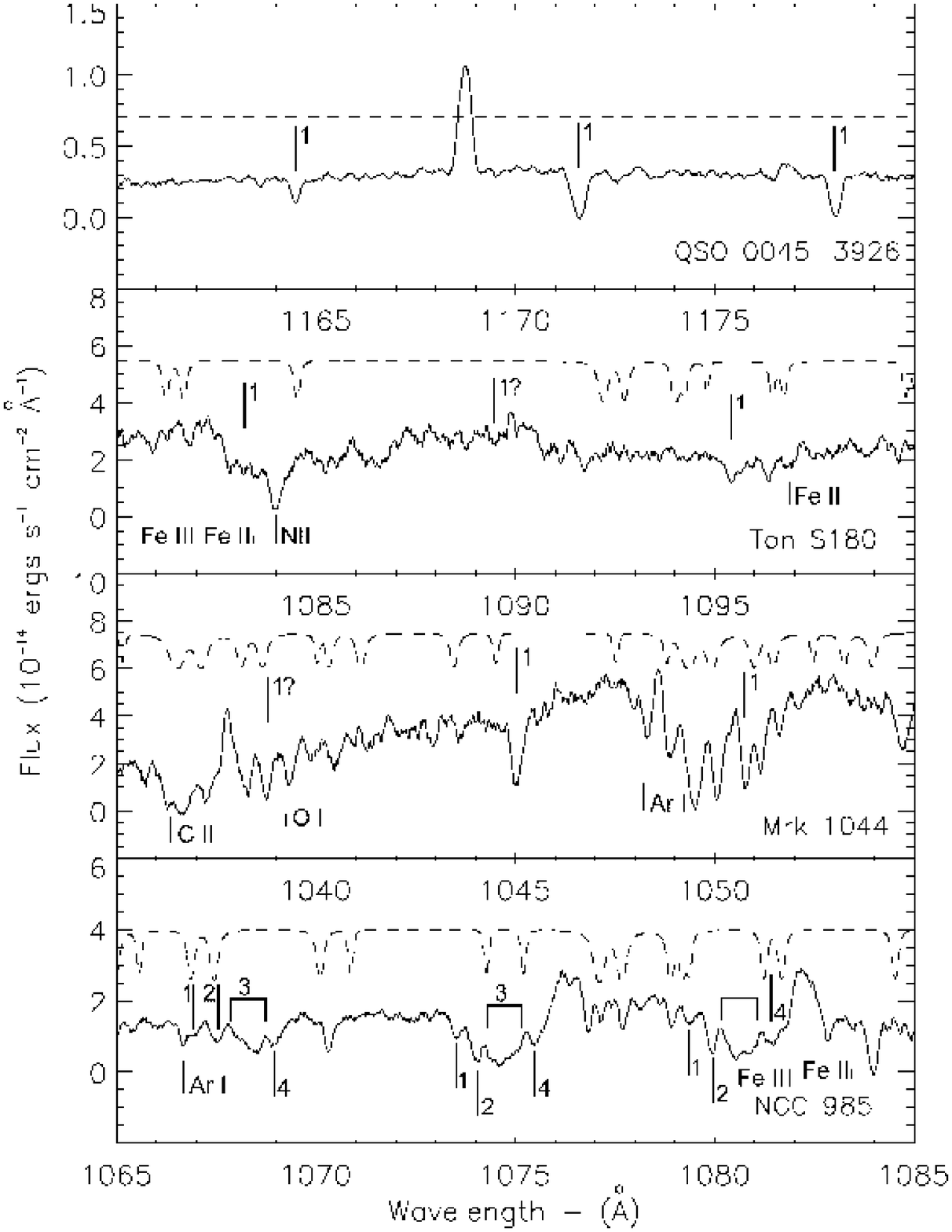}
\\Fig.~2.
\end{figure}

\clearpage
\begin{figure}
\plotone{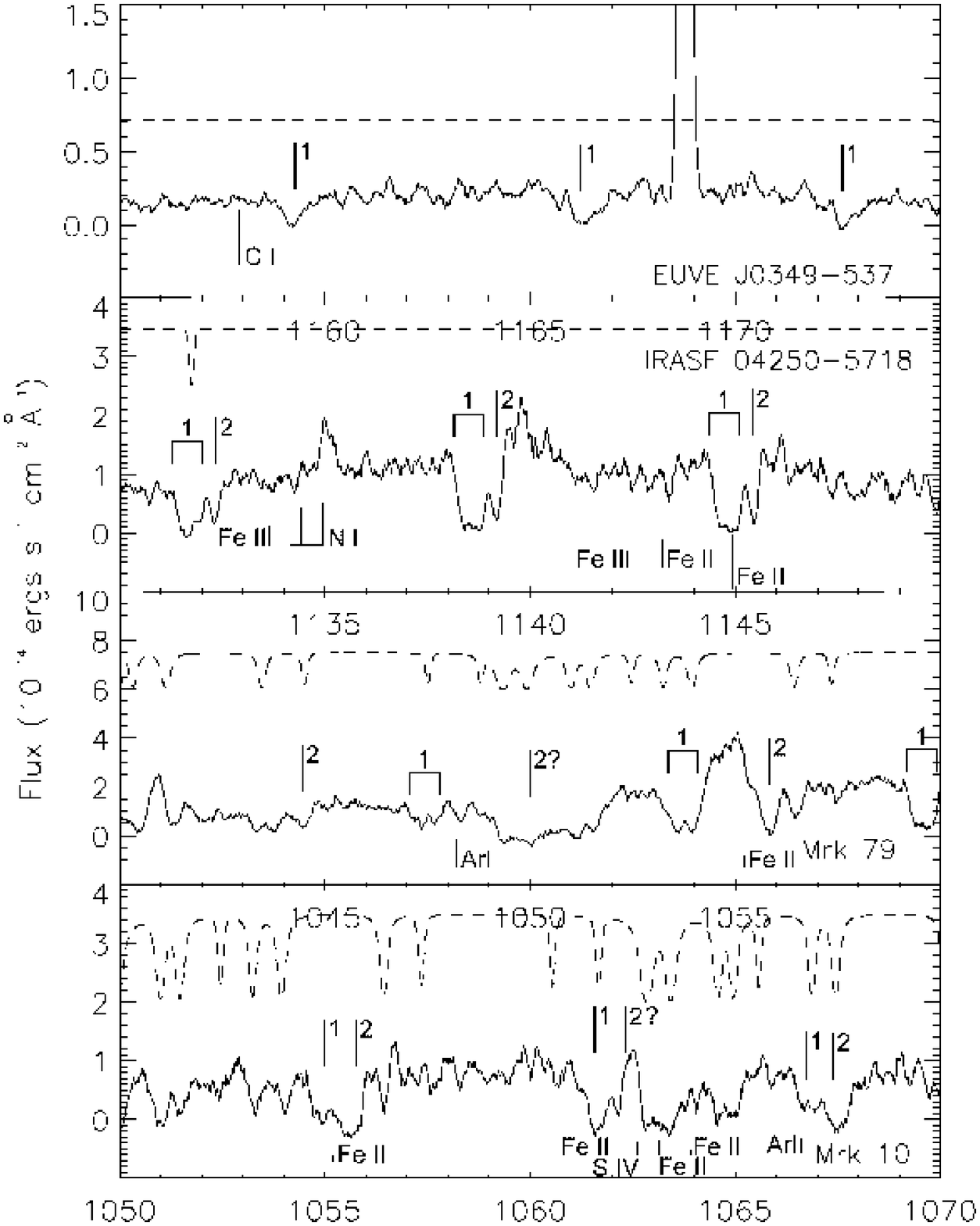}
\\Fig.~2.
\end{figure}

\clearpage
\begin{figure}
\plotone{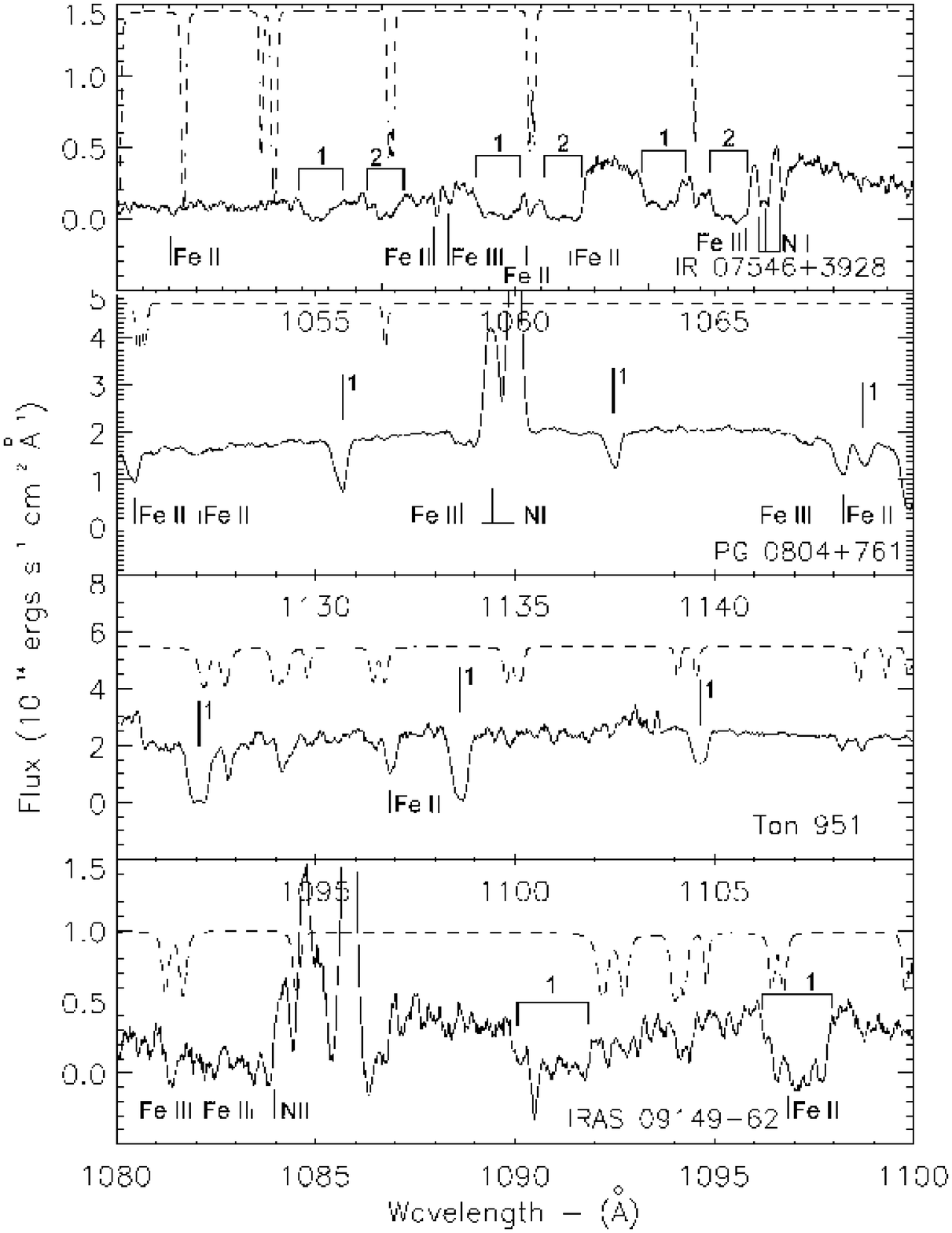}
\\Fig.~2.
\end{figure}

\clearpage
\begin{figure}
\plotone{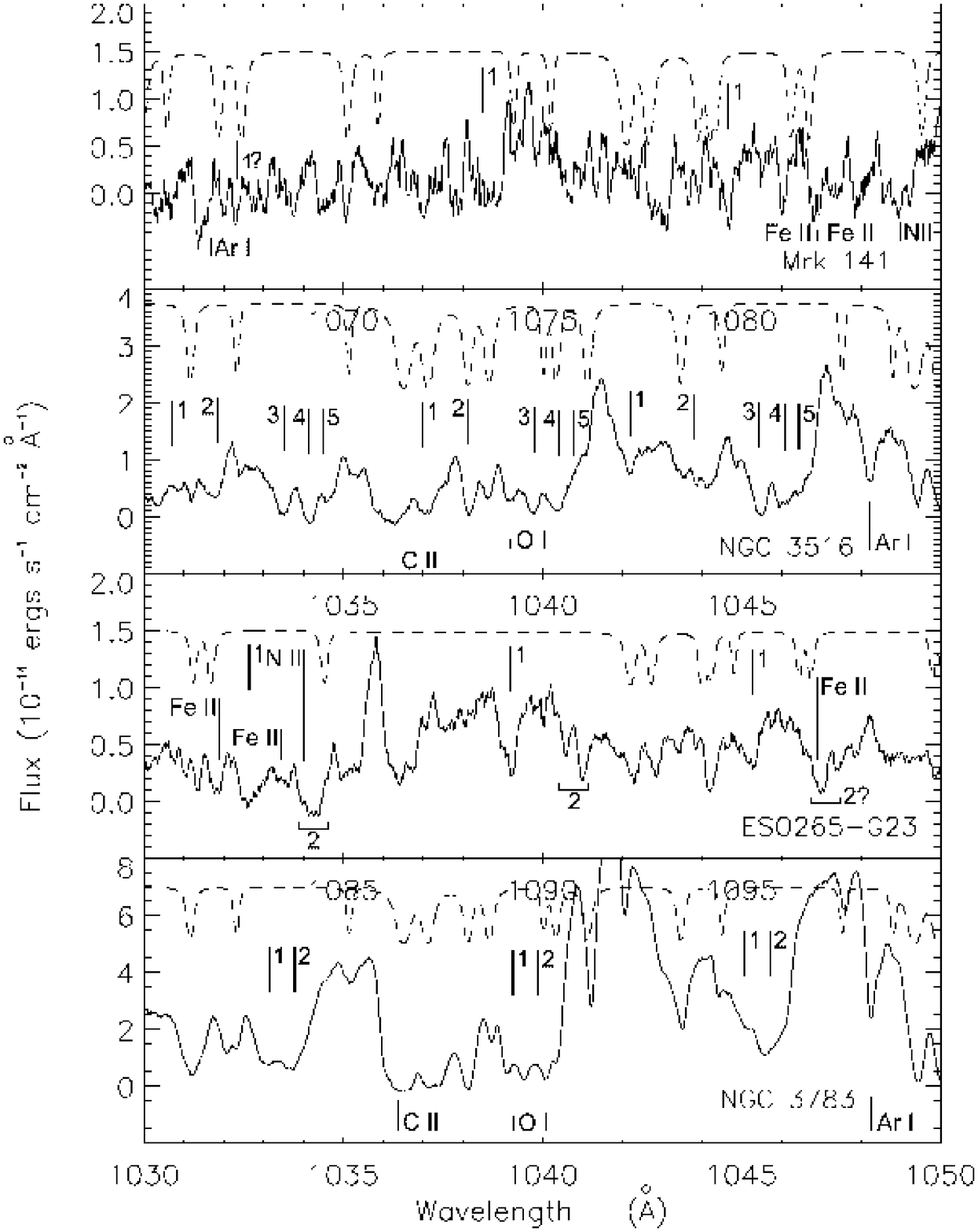}
\\Fig.~2.
\end{figure}

\clearpage
\begin{figure}
\plotone{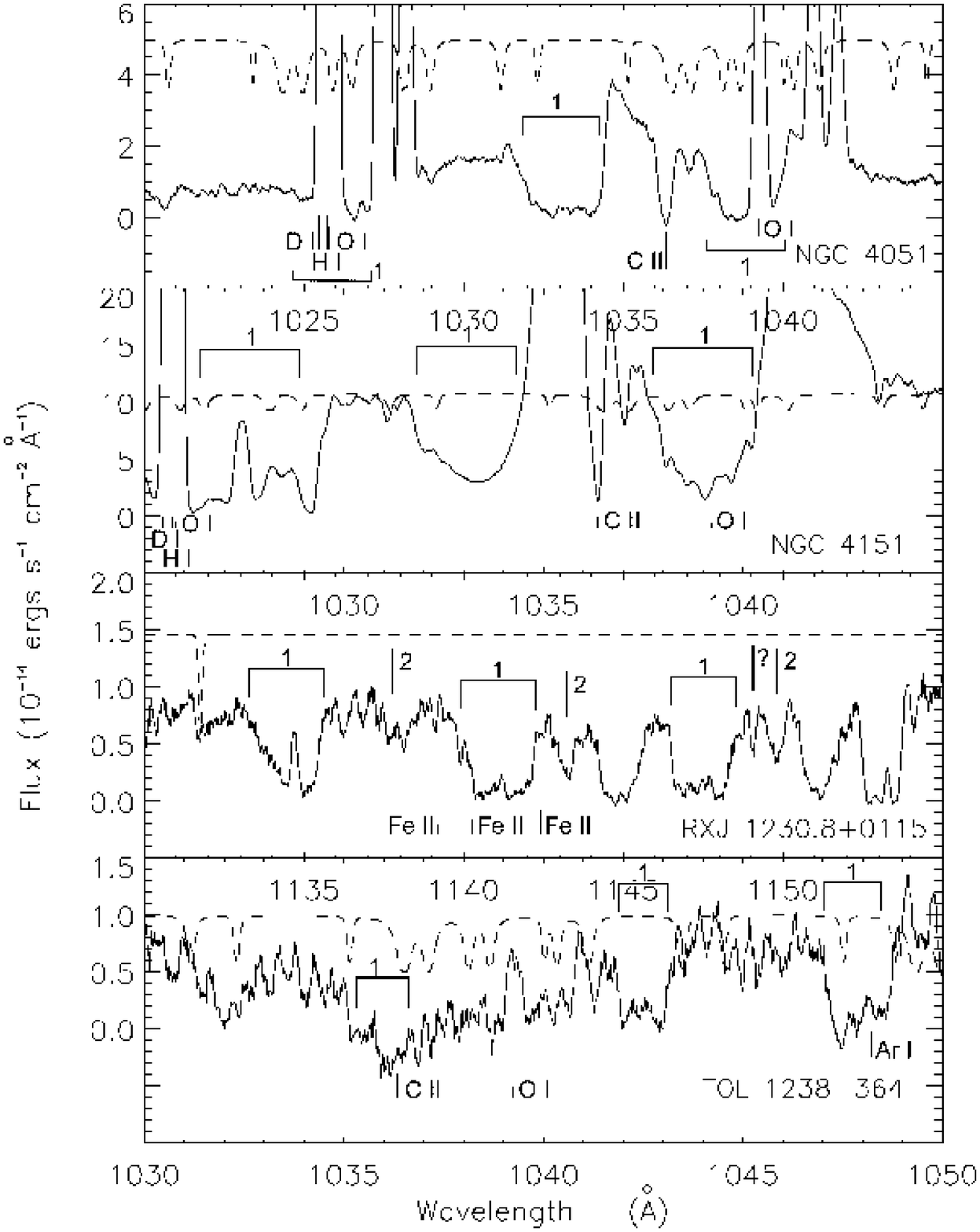}
\\Fig.~2.
\end{figure}

\clearpage
\begin{figure}
\plotone{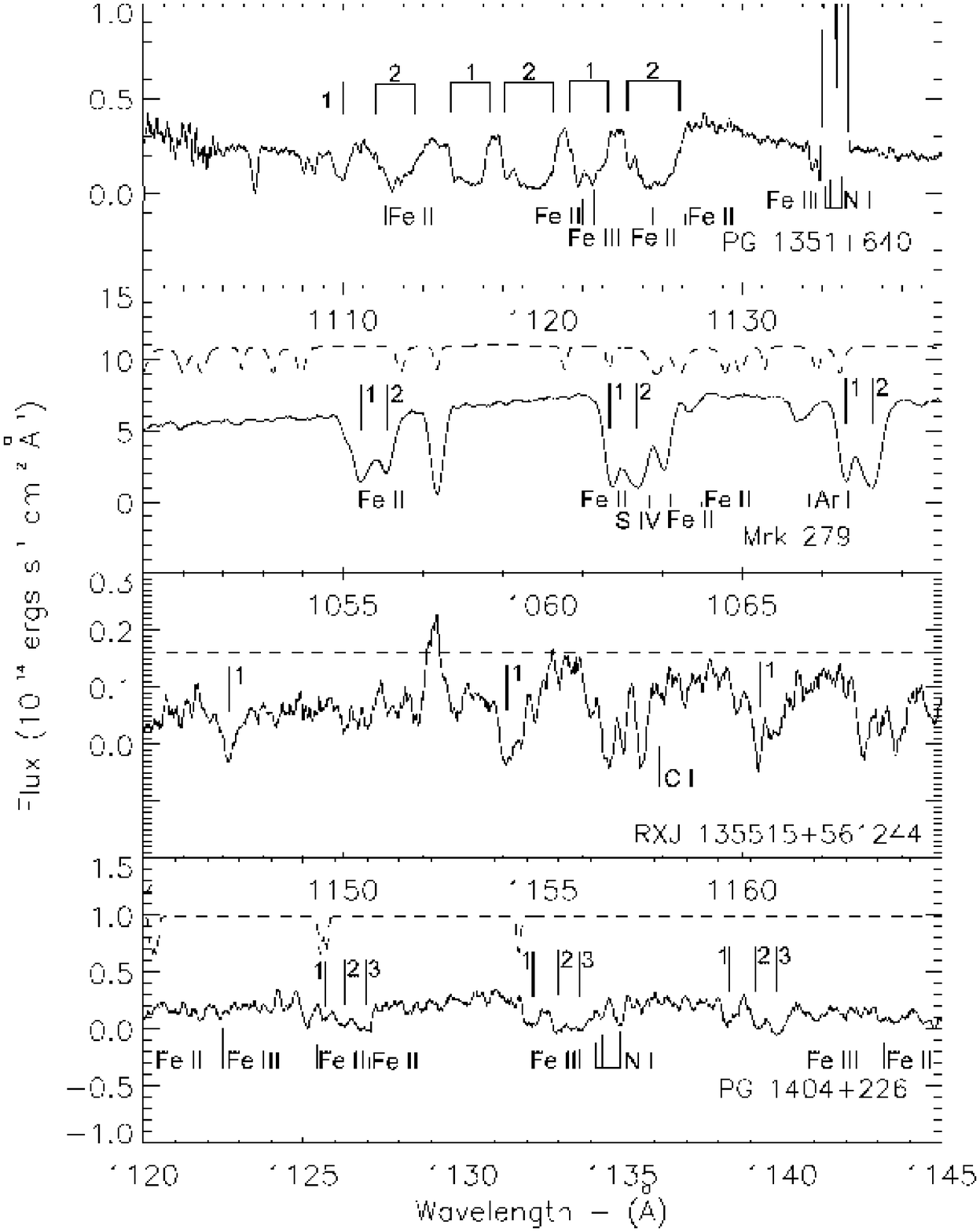}
\\Fig.~2.
\end{figure}

\clearpage
\begin{figure}
\plotone{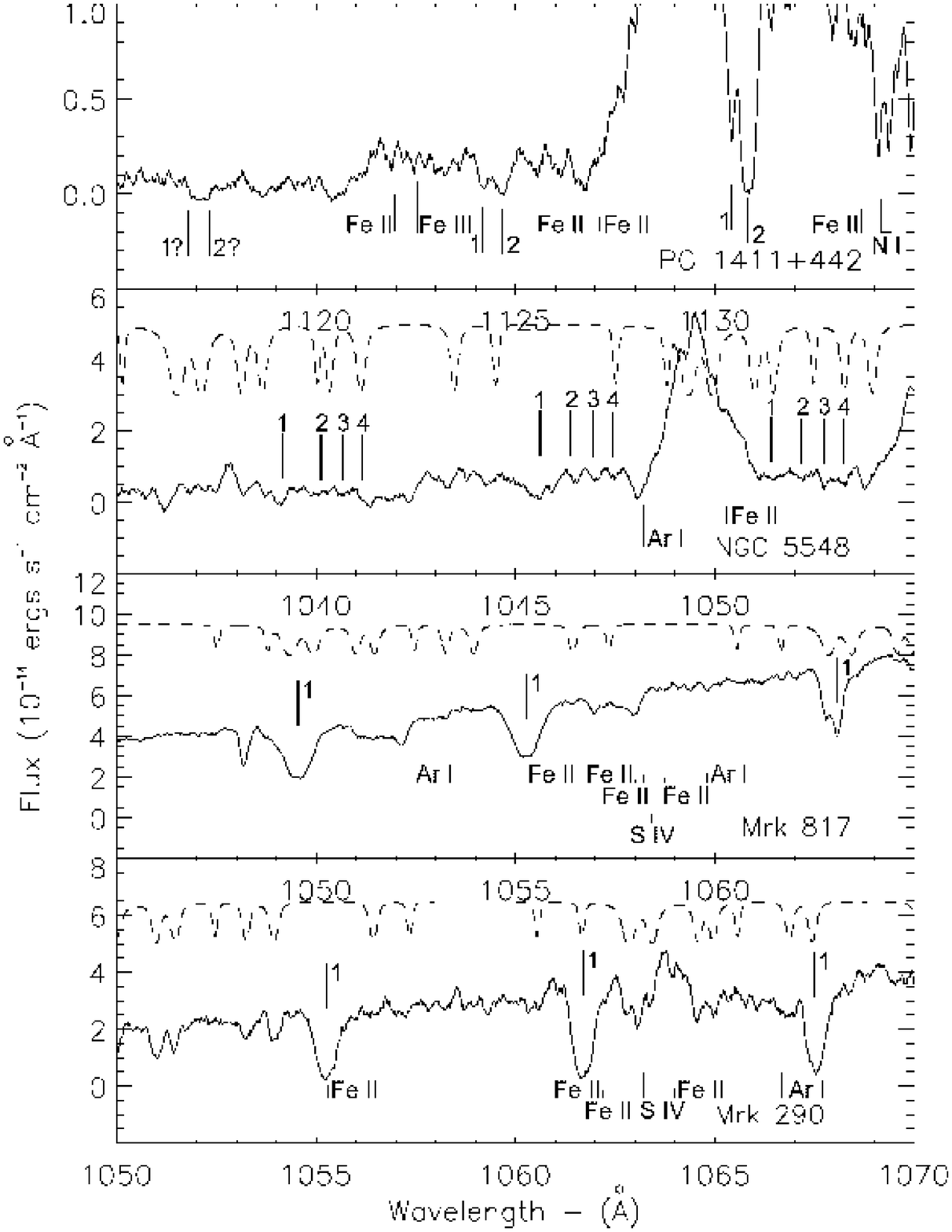}
\\Fig.~2.
\end{figure}

\clearpage
\begin{figure}
\plotone{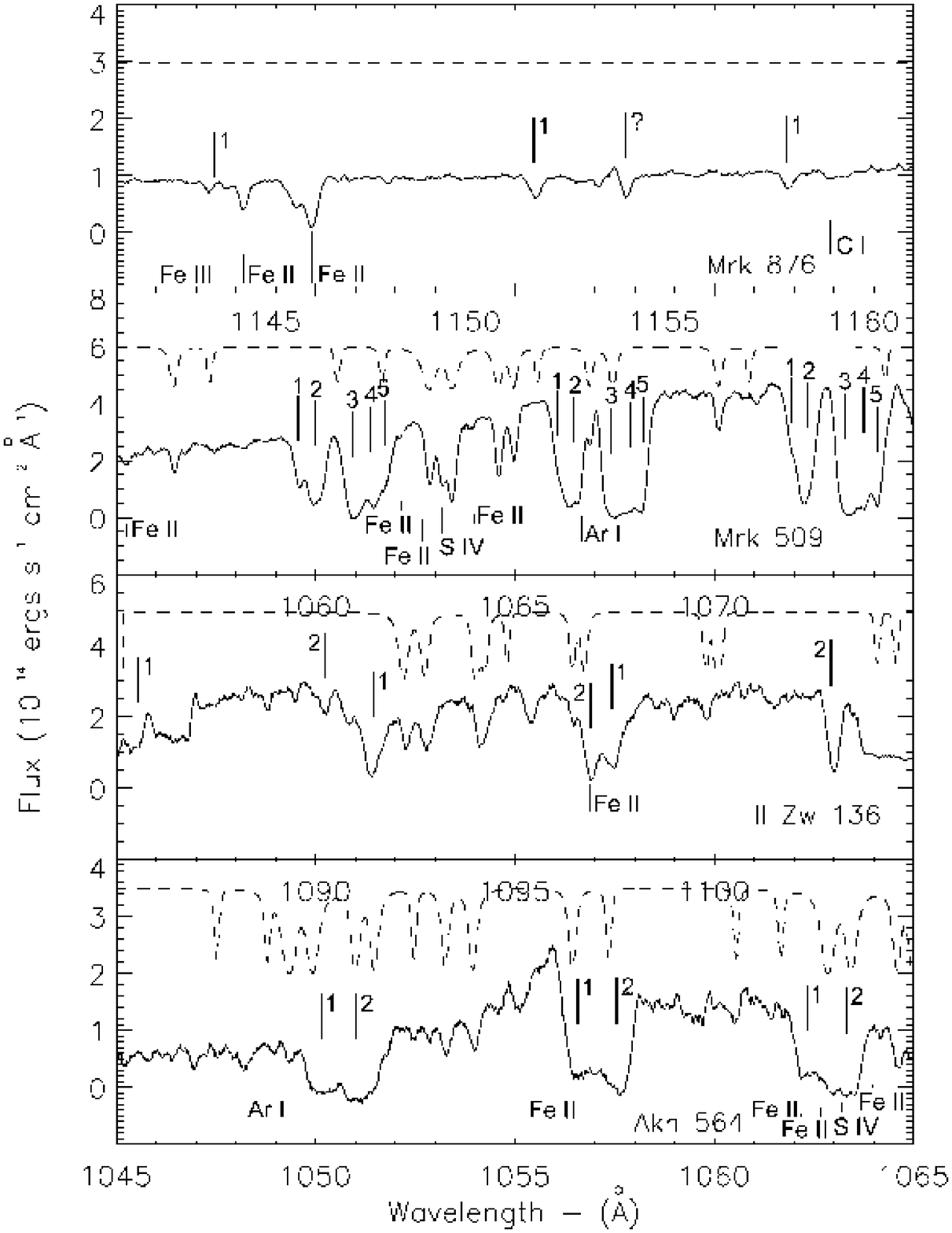}
\\Fig.~2.
\end{figure}

\clearpage
\begin{figure}
\plotone{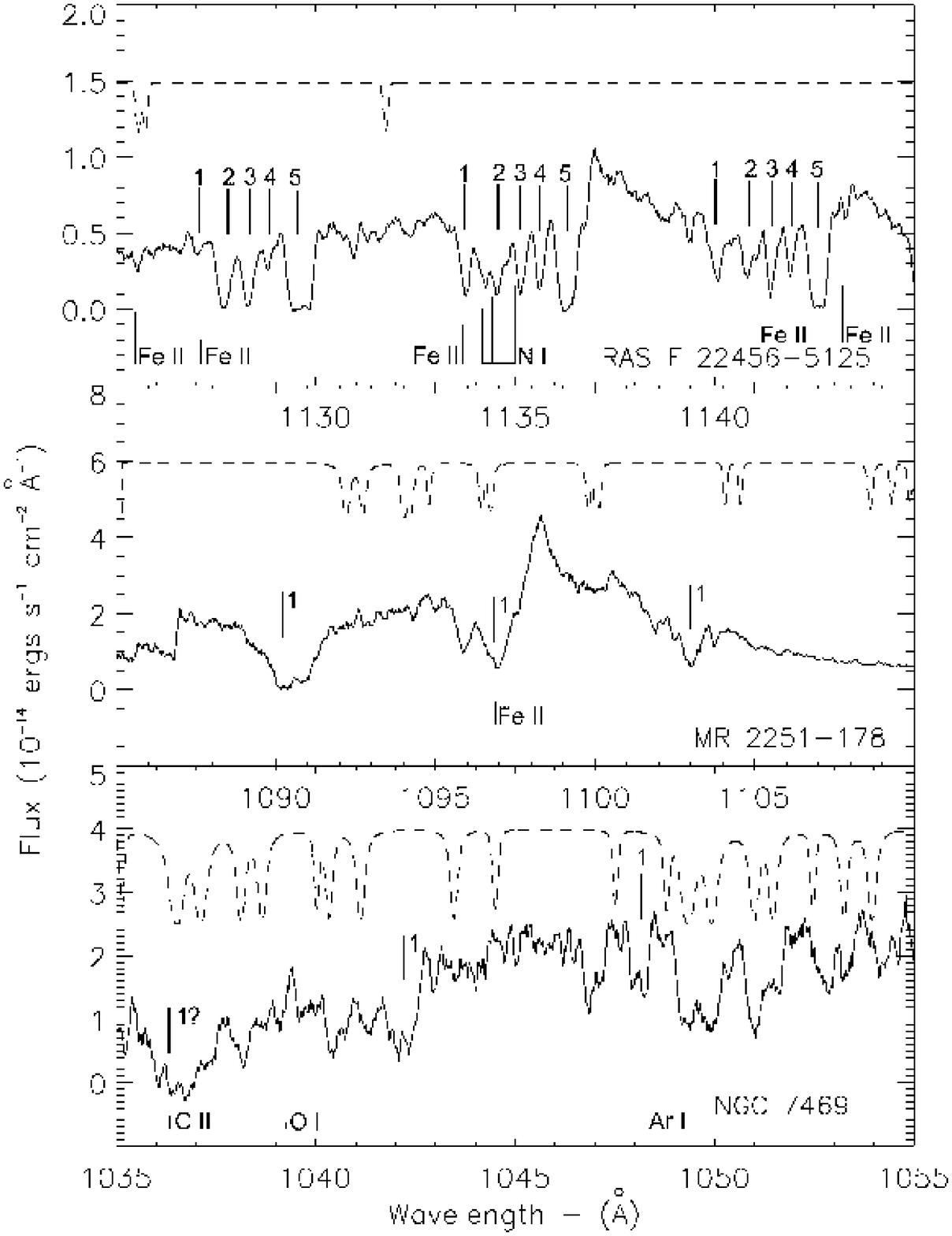}
\\Fig.~2.
\end{figure}

%\clearpage
%\begin{figure}
%\plotone{fg2j.eps}
%\\Fig.~2.
%\end{figure}

\newpage
\begin{deluxetable}{lccclcclc}
\rotate
\tablecolumns{9}
\footnotesize
\tablecaption{AGN Included in the Survey}
\tablewidth{1.2\textwidth}
\tabletypesize{\scriptsize}
\tablehead{
\colhead{Object} &
\colhead{RA} &
\colhead{Dec} &
\colhead{z} &
\colhead{Class$^b$} &
\colhead{S/N} &
\colhead{Observation ID} &
\colhead{Observation Date} &
\colhead{Exposure Time (s)}
}
\startdata

MRK335   & 00 06 19.53   & +20 12 10.3   & 0.026 & Sey 1.2 & 54.9 & P1010204000  & 11/21/2000 & 53391 \\
 &  &  &  &  &  & P1010203000   & 12/4/1999 & 45894 \\
%WPVS007  & 00 39 15.81  & -51 17 1.9  & 0.029 & Sey 1 & 5.1 & D8060201000  & 11/6/2003 & 48135 \\
QSO0045+3926  & 00 48 18.90  & +39 41 12.0  & 0.134 & Sey 1 & 9.6 & D1310101000  & 10/8/2003 & 42658 \\
 &  &  &  &  &  & D1310105000  & 11/25/2004 & 25452 \\
 &  &  &  &  &  & D1310106000  & 11/26/2004 & 27487 \\
 &  &  &  &  &  & D1310104000  & 12/9/2003 & 40728 \\
 &  &  &  &  &  & D1310107000  & 11/27/2004 & 25058 \\
 &  &  &  &  &  & D1310102000  & 10/10/2003 & 34371 \\
 &  &  &  &  &  & Z0020401000  & 11/25/2000 & 8561 \\
 &  &  &  &  &  & D1310103000  & 10/11/2003 & 7272 \\
I ZW 1  & 00 53 34.90  & +12 41 36.0  & 0.061 & Sey 1 & 3.8 & P1110101000  & 12/3/1999 & 13584 \\
 &  &  &  &  &  & P1110102000  & 11/20/2000 & 25189 \\
TONS180  & 00 57 19.95  & -22 22 59.3  & 0.062 & Sey 1.2 & 14.0 & P1010502000  & 12/12/1999 & 15420 \\
 &  &  &  &  &  & D0280101000  & 7/13/2004 & 8955 \\
MRK352  & 00 59 53.28  & +31 49 36.7  & 0.015 & Sey 1 & 3.7 & P1070201000  & 10/11/1999 & 16644 \\
RXJ010027-511346  & 01 00 27.06  & -51 13 54.8  & 0.063 & Sey 1 & 2.5 & D8060301000  & 9/11/2003 & 6321 \\
 &  &  &  &  &  & E8970201000  & 8/22/2004 & 16845 \\
TONS210  & 01 21 51.56  & -28 20 57.3  & 0.116 & Sey 1 & 23.0 & P1070301000  & 10/21/1999 & 14023 \\
 &  &  &  &  &  & P1070302000  & 8/10/2001 & 41023 \\
FAIRALL9  & 01 23 46.04  & -58 48 23.8  & 0.047 & Sey 1 & 12.54 & P1010601000  & 7/3/2000 & 34896 \\
MRK1044  & 02 30 5.45  & -08 59 52.6  & 0.016 & Sey 1 & 7.3 & D0410101000  & 1/1/2004 & 12608 \\
NGC985  & 02 34 37.77  & -08 47 15.6  & 0.043 & Sey 1 & 29.4 & P1010903000  & 12/24/1999 & 49968 \\
ESO31-8  & 03 07 35.30  & -72 50 6.2  & 0.028 & Sey 1 & 3.3 & D9110201000  & 7/13/2003 & 15915 \\
EUVEJ0349-537  & 03 49 28.50  & -53 44 47.0  & 0.130 & Sey 1 & 2.9 & E8970301000  & 8/20/2004 & 27732 \\
IRASF04250-5718  & 04 26 0.83  & -57 12 0.4  & 0.104 & Sey 1 & 2.1 & D8080801000  & 9/4/2003 & 4895 \\
FAIR303  & 04 30 40.02  & -53 36 55.9  & 0.040 & Sey 1 & 3.1 & D8060702000  & 10/31/2003 & 16728 \\
MRK618  & 04 36 22.25  & -10 22 33.9  & 0.036 & Sey 1 & 3.2 & P1070901000  & 10/21/1999 & 2691 \\
 &  &  &  &  &  & P1070902000  & 12/7/2000 & 6293 \\
AKN120  & 05 16 11.42  & -00 08 59.4  & 0.033 & Sey 1 & 12.2 & P1011203000  & 12/31/2000 & 25617 \\
 &  &  &  &  &  & P1011201000  & 11/1/2000 & 8448 \\
 &  &  &  &  &  & P1011202000  & 11/3/2000 & 20942 \\
PKS0558-504  & 05 59 47.40  & -50 26 52.0  & 0.137 & NL Sey 1 & 18.4 & C1490601000  & 11/7/2002 & 48480 \\
 &  &  &  &  &  & P1011504000  & 12/10/1999 & 47271 \\
IRASL06229-6434  & 06 23 9.10  & -64 36 24.0  & 0.129 & Sey 1 & 2.9 & D9030304000  & 12/17/2003 & 14494 \\
 &  &  &  &  &  & U1071002000  & 7/19/2006 & 18569 \\
 &  &  &  &  &  & D9030301000  & 12/14/2003 & 51104 \\
 &  &  &  &  &  & U1071001000  & 7/17/2006 & 11284 \\
 &  &  &  &  &  & D9030302000  & 12/16/2003 & 23363 \\
 &  &  &  &  &  & D9030303000  & 12/17/2003 & 11122 \\
VIIZW118  & 07 07 13.10  & +64 35 58.8  & 0.080 & Sey 1 & 11.6 & P1011606000  & 1/1/2000 & 9805 \\
 &  &  &  &  &  & P1011604000  & 10/6/1999 & 77568 \\
 &  &  &  &  &  & P1011605000  & 11/29/1999 & 26075 \\
 &  &  &  &  &  & P1011601000  & 10/2/1999 & 47515 \\
 &  &  &  &  &  & S6011301000  & 2/8/2002 & 47363 \\
 &  &  &  &  &  & P1011603000  & 10/2/1999 & 8957 \\
1H0707-495  & 07 08 41.50  & -49 33 5.8  & 0.041 & Sey 1 & 12.8 & B1050101000  & 11/16/2001 & 19989 \\
 &  &  &  &  &  & E1190101000  & 12/10/2004 & 50252 \\
 &  &  &  &  &  & B1050102000  & 3/6/2003 & 24260 \\
 &  &  &  &  &  & B1050103000  & 3/6/2003 & 22576 \\
MRK9  & 07 36 57.02  & +58 46 13.4  & 0.040 & Sey 1.5 & 5.3 & P1071102000  & 1/8/2000 & 14278 \\
 &  &  &  &  &  & P1071101000  & 11/29/1999 & 11829 \\
 &  &  &  &  &  & P1071103000  & 2/22/2000 & 10664 \\
 &  &  &  &  &  & S6011601000  & 2/11/2002 & 23555 \\
MRK79  & 07 42 32.80  & +49 48 34.9  & 0.022 & Sey 1.2 & 3.0 & P1011702000  & 1/14/2000 & 11688 \\
 &  &  &  &  &  & P1011701000  & 11/30/1999 & 3147 \\
 &  &  &  &  &  & P1011703000  & 2/22/2000 & 12498 \\
MRK10  & 07 47 29.10  & +60 56 1.0  & 0.029 & Sey 1.2 & 2.7 & Z9072801000  & 2/2/2003 & 22699 \\
IR07546+3928  & 07 58 0.05  & +39 20 29.1  & 0.096 & Sey 1.5 & 6.7 & P1071201000  & 3/9/2000 & 15290 \\
 &  &  &  &  &  & S6011801000  & 2/11/2002 & 27410 \\
 &  &  &  &  &  & E0870101000  & 11/7/2004 & 53151 \\
PG0804+761  & 08 10 58.46  & +76 02 41.9  & 0.100 & Sey 1 & 28.3 & S6011002000  & 2/9/2002 & 36886 \\
 &  &  &  &  &  & S6011001000  & 2/5/2002 & 58506 \\
 &  &  &  &  &  & P1011901000  & 10/5/1999 & 42715 \\
 &  &  &  &  &  & P1011903000  & 1/4/2000 & 21128 \\
UGC4305  & 08 19 12.90  & +70 43 6.0  & 0.001 & Sey 1 & 11.1 & F0270104000  & 12/20/2005 & 2210 \\
 &  &  &  &  &  & F0270102000  & 12/18/2005 & 28130 \\
 &  &  &  &  &  & F0270103000  & 12/19/2005 & 18821 \\
PG0838+770  & 08 44 45.26  & +76 53 10.0  & 0.132 & Sey 1 & 2.62 & G0200101000  & 2/10/2006 & 5942 \\
 &  &  &  &  &  & G0200104000  & 2/13/2006 & 35047 \\
Ton951  & 08 47 42.60  & +34 45 4.7  & 0.064 & Sey 1 & 7.515 & D0280304000  & 3/15/2004 & 10874 \\
 &  &  &  &  &  & P1012002000  & 2/20/2000 & 30962 \\
 &  &  &  &  &  & D0280303000  & 3/15/2004 12:06 & 12774 \\
 &  &  &  &  &  & D0280302000  & 3/15/2004 3:50 & 12713 \\
 &  &  &  &  &  & D0280301000  & 3/14/2004 14:28 & 7935 \\
IRAS09149-62  & 09 16 9.41  & -62 19 29.5  & 0.057 & Sey 1 & 6.45 & S7011003000  & 3/31/2005 & 14520 \\
 &  &  &  &  &  & S7011002000  & 3/30/2005 & 13648 \\
 &  &  &  &  &  & A0020503000  & 2/6/2000 & 13157 \\
 &  &  &  &  &  & U1072201000  & 1/19/2006 & 7554 \\
MRK110  & 09 25 12.87  & +52 17 10.7  & 0.035 & Sey 1 & 2.52 & P1071302000  & 2/11/2001 & 10791 \\
TON1187  & 10 13 3.21  & +35 51 22.2  & 0.079 & Sey 1 & 2.90 & P1071502000  & 1/13/2000 & 7994 \\
PG1011-040  & 10 14 20.58  & -04 18 41.2  & 0.058 & Sey 1 & 35.78 & B0790101000  & 5/16/2001 & 85197 \\
MKN141  & 10 19 12.59  & +63 58 2.7  & 0.042 & Sey 1.5 & 1.73 & D8061001000  & 3/21/2004 & 15768 \\
MKN142  & 10 25 31.28  & +51 40 34.9  & 0.045 & Sey 1 & 8.53 & D8061101000  & 3/23/2003 & 20103 \\
KUV-1031+398  & 10 34 38.61  & +39 38 28.4  & 0.042 & Sey 1 & 7.08 & A0990101000  & 4/30/2000 & 53185 \\
NGC3516 & 11 06 47.55  & +72 34 6.9  & 0.009 & Sey 1.5 & 1.91$^a$ & P2110103000  & 1/28/2003 & 16753 \\
 &  &  &  &  &  & G9170101000  & 2/9/2006 & 28718 \\
 &  &  &  &  &  & P1110404000  & 4/17/2000 & 16335 \\
 &  &  &  &  &  & P2110102000  & 2/14/2002 & 20702 \\
 &  &  &  &  &  & P2110104000  & 3/29/2003 & 16125 \\
 &  &  &  &  &  & G9170102000 & 1/23/2007 & 16877 \\
ESO265-G23  & 11 20 47.89  & -43 15 50.6  & 0.056 & Sey 1 & 3.2 & A1210409000  & 4/29/2002 & 6428 \\
 &  &  &  &  &  & A1210405000  & 3/9/2001 & 4863 \\
 &  &  &  &  &  & A1210408000  & 4/28/2002 & 11487 \\
 &  &  &  &  &  & A1210407000  & 3/7/2002 & 21716 \\
 &  &  &  &  &  & A1210406000  & 2/28/2002 & 8381 \\
 &  &  &  &  &  & A1210404000  & 5/28/2000 & 1782 \\
MRK734  & 11 21 47.11  & +11 44 18.5  & 0.050 & Sey 1.2 & 3.1 & P1071702000  & 4/16/2001 & 4587 \\
NGC3783  & 11 39 1.78  & -37 44 18.5  & 0.010 & Sey 1 & 21.6 & B1070103000  & 3/11/2001 & 27221 \\
 &  &  &  &  &  & B1070104000  & 3/30/2001 & 25180 \\
 &  &  &  &  &  & B1070105000  & 6/27/2001 & 27483 \\
 &  &  &  &  &  & B1070106000  & 3/7/2001 & 25541 \\
 &  &  &  &  &  & P1013301000  & 2/2/2000 & 37003 \\
 &  &  &  &  &  & E0310101000  & 5/5/2004 & 23286 \\
 &  &  &  &  &  & B1070102000  & 2/28/2001 & 26645 \\
IR1143-1810  & 11 45 40.48  & -18 27 15.3  & 0.033 & Sey 1 & 8.51 & P1071901000  & 5/29/2000 & 7243 \\
NGC4051  & 12 03 9.61  & +44 31 52.8  & 0.002 & Sey 1.5 & 7.26 & B0620201000  & 3/29/2002 & 28659 \\
 &  &  &  &  &  & C0190102000  & 3/19/2003 & 28594 \\
 &  &  &  &  &  & C0190101000  & 1/18/2003 & 13933 \\
NGC4151  & 12 10 32.60  & +39 24 21.0  & 0.003 & Sey 1.5 & 61.79 & P2110201000  & 4/8/2001 & 13612 \\
 &  &  &  &  &  & P2110202000  & 6/1/2002 & 5953 \\
 &  &  &  &  &  & C0920101000  & 5/28/2002 & 48892 \\
 &  &  &  &  &  & P1110505000  & 3/5/2000 & 21522 \\
PG1211+143  & 12 14 17.61  & +14 03 12.7  & 0.081 & Sey 1 & 40.02 & P1072001000  & 4/25/2000 & 52274 \\
MRK205  & 12 21 44.04  & +75 18 38.3  & 0.071 & Sey 1 & 24.71 & S6010801000  & 2/2/2002 & 16027 \\
 &  &  &  &  &  & D0540101000  & 11/13/2003 & 19618 \\
 &  &  &  &  &  & D0540103000  & 11/17/2003 & 19131 \\
 &  &  &  &  &  & Q1060203000  & 12/29/1999 & 37051 \\
 &  &  &  &  &  & D0540102000  & 11/14/2003 & 114015 \\
NGC4395  & 12 25 48.92  & +33 32 48.4  & 0.001 & Sey 1 & 4.39 & C0860101000  & 2/25/2003 & 36484 \\
RXJ1230.8+0115  & 12 30 50.00  & +01 15 22.7  & 0.117 & Unk$^c$ & 2.79 & P1019001000  & 6/20/2000 & 4031 \\
PG1229+204  & 12 32 3.62  & +20 09 29.4  & 0.063 & Sey 1 & 2.62 & P1072301000  & 2/5/2001 & 6267 \\
TOL1238-364  & 12 40 52.90  & -36 45 21.1  & 0.011 & Sey 2? & 4.67 & D0100101000  & 6/13/2003 & 17912 \\
PG1351+640  & 13 53 15.80  & +63 45 45.0  & 0.088 & Sey 1 & 15.42 & S6010701000  & 2/1/2002 & 48620 \\
 &  &  &  &  &  & P1072501000  & 1/18/2000 & 70134 \\
MRK279  & 13 53 3.52  & +69 18 29.7  & 0.030 & Sey 1.5 & 53.24 & P1080304000  & 1/11/2000 & 30747 \\
 &  &  &  &  &  & F3250103000  & 12/7/2005 & 2289 \\
 &  &  &  &  &  & C0900201000  & 5/18/2002 & 41708 \\
 &  &  &  &  &  & D1540101000  & 5/12/2003 & 91040 \\
 &  &  &  &  &  & F3250104000  & 12/8/2005 & 3187 \\
 &  &  &  &  &  & F3250106000  & 2/3/2006 & 3703 \\
 &  &  &  &  &  & P1080303000  & 12/28/1999 & 60338 \\
RXJ135515+561244  & 13 55 16.55  & +56 12 44.6  & 0.122 & Sey 1 & 1.78 & D8061601000  & 3/13/2003 & 47223 \\
PG1404+226  & 14 06 22.15  & +22 23 42.8  & 0.098 & Sey 1 & 1.86 & P2100401000  & 6/11/2001 & 11489 \\
PG1411+442  & 14 13 48.32  & +44 00 13.1  & 0.090 & Sey 1 & 2.64 & A0601010000  & 5/11/2000 & 7360 \\
PG1415+451  & 14 17 0.84  & +44 56 6.0  & 0.114 & Sey 1 & 2.47 & A0601111000  & 5/10/2000 & 12285 \\
NGC5548  & 14 17 59.91  & +25 08 12.6  & 0.017 & Sey 1.5 & 11.11 & D1550102000  & 2/11/2004 & 7757 \\
 &  &  &  &  &  & P1014601000  & 6/7/2000 & 25932 \\
 &  &  &  &  &  & D1550101000  & 2/10/2004 & 22229 \\
MRK1383  & 14 29 6.60  & +01 17 6.6  & 0.086 & Sey 1 & 22.27 & P1014801000  & 2/18/2000 & 25051 \\
MRK817  & 14 36 22.09  & +58 47 39.5  & 0.031 & Sey 1.5 & 49.78 & P1080404000  & 2/18/2001 & 86013 \\
 &  &  &  &  &  & P1080402000  & 2/18/2000 & 12911 \\
 &  &  &  &  &  & P1080401000  & 2/17/2000 & 12119 \\
 &  &  &  &  &  & P1080403000  & 12/23/2000 & 71804 \\
MRK477  & 14 40 38.06  & +53 30 15.7  & 0.038 & Sey 1 & 7.25 & P1110808000  & 5/8/2001 & 11289 \\
MRK478  & 14 42 7.46  & +35 26 22.9  & 0.079 & Sey 1 & 1.86 & P1110909000  & 1/29/2001 & 14118 \\
MRK290  & 15 35 52.38  & +57 54 9.3  & 0.030 & Sey 1 & 5.53 & D0760101000  & 6/28/2003 & 9239 \\
 &  &  &  &  &  & D0760102000  & 2/27/2004 & 46032 \\
 &  &  &  &  &  & P1072901000  & 3/16/2000 & 12769 \\
 &  &  &  &  &  & E0840101000  & 6/15/2004 & 11931 \\
MRK876  & 16 13 57.21  & +65 43 9.7  & 0.129 & Sey 1 & 37.02 & P1073101000  & 10/16/1999 & 52659 \\
 &  &  &  &  &  & D0280201000  & 5/16/2003 & 11786 \\
 &  &  &  &  &  & D0280203000  & 2/14/2004 & 73334 \\
PG1626+554  & 16 27 56.09  & +55 22 32.0  & 0.133 & Sey 1 & 2.15 & P1073201000  & 2/18/2000 & 9754 \\
 &  &  &  &  &  & C0370101000  & 5/20/2002 & 91237 \\
%MRK501  & 16 53 52.22  & +39 45 36.6  & 0.034 & BLLAC & 9.01 & C0810101000  & 4/26/2004 & 20141 \\
MRK506  & 17 22 39.92  & +30 52 53.1  & 0.043 & Sey 1 & 4.34 & P1073401000  & 6/8/2000 & 10430 \\
3C382  & 18 35 3.38  & +32 41 47.0  & 0.058 & Sey 1 & 3.15 & P1073701000  & 6/9/2000 & 11297 \\
PKS2005-489  & 20 09 25.39  & -48 49 53.7  & 0.071 & QSO & 6.97 & C1490301000  & 4/12/2002 & 24726 \\
 &  &  &  &  &  & C1490302000  & 6/4/2002 & 13422 \\
MRK509  & 20 44 9.74  & -10 43 24.7  & 0.034 & Sey 1.2 & 42.83 & P1080601000  & 9/5/2000 & 60656 \\
 &  &  &  &  &  & X0170102000  & 11/6/1999 & 33410 \\
 &  &  &  &  &  & X0170101000  & 11/2/1999 & 20004 \\
%1H2107-097 & 21 09 9.96  & -09 40 15.0  & 0.027 & Sey 1.2 & 1.14  & A1210303000  & 6/17/2001 & 5973 \\
IIZW136  & 21 32 27.83  & +10 08 19.4  & 0.063 & Sey 1 & 11.55 & P1018301000  & 11/14/2000 & 22629 \\
 &  &  &  &  &  & P1018302000  & 5/27/2004 & 11497 \\
 &  &  &  &  &  & P1018303000  & 5/28/2004 & 8067 \\
 &  &  &  &  &  & P1018304000  & 11/1/2004 & 21872 \\
MRK304  & 22 17 12.28  & +14 14 20.9  & 0.066 & RQQ & 3.05 & P1073901000  & 7/16/2000 & 12387 \\
AKN564  & 22 42 39.34  & +29 43 31.3  & 0.025 & Sey 1.8 & 6.19 & B0620101000  & 6/29/2001 & 55515 \\
IRASF22456-5125  & 22 48 41.00  & -51 09 54.0  & 0.100 & Sey 1 & 6.82 & Z9073902000  & 9/24/2002 & 31301 \\
 &  &  &  &  &  & Z9073901000  & 9/24/2002 & 5534 \\
MR 2251-178  & 22 54 5.80  & -17 34 55.0  & 0.064 & Sey 1 & 16.13 & P1111010000  & 6/20/2001 & 54113 \\
NGC7469  & 23 03 15.62  & +08 52 25.6  & 0.016 & Sey 1.2 & 2.22 & C0900101000  & 12/13/2002 & 3593 \\
 &  &  &  &  &  & C0900102000  & 12/14/2002 & 3352 \\
 &  &  &  &  &  & P1074101000  & 6/28/2000 & 13217 \\

\enddata
\normalsize
\tablenotetext{a}{Reduced from a possible S/N of 3.34.}
\tablenotetext{b}{AGN types listed in the NASA Extragalactic Database}
\tablenotetext{c}{No AGN type listed}
\end{deluxetable}
\thispagestyle{empty}

\begin{deluxetable}{ll}
\tablecolumns{6}
\footnotesize
\tablecaption{Objects with Intrinsic Absorption in the Survey}
%\tablewidth{Opt}
\tablehead{
\colhead{Object} &
\colhead{Component Velocities (km s$^-$$^1$)}
}
\startdata

QSO0045+3926 & 1) +340 \\
TONS180 & 1) -1800 \\
MRK1044 & 1) -1100, 2) -270 \\
NGC985 & 1) -700, 2) -410, 3) -290 \\
EUVEJ0349-537 & 1) 0 \\
IRASF04250-5718 & 1) -200, 2) -100, 3) -20 \\
MRK79 & 1) -1400, 2) -320 \\
MRK10 & 1) -150, 2) -200 \\
IR07546+3928 & 1) -1800, 2) -1200\\
PG0804+761 & 1) +700 \\
TON951 & 1) +160 \\
IRAS09149-62 & 1) 0 \\
MKN141 & 1) -600 \\
NGC3516 & 1), -1320 2), -830 3), -370 4), -180 5) -60\\
ESO265-G23 & 1) -150, 2) +400 \\
NGC3783 & 1) -700 \\
NGC4051 & 1) -300 \\
NGC4151 & 1) -500 \\
RXJ1230.8+0115 & 1) -3000, 2) -2000, 3) +400 \\
TOL1238-364 & 1) -200 \\
PG1351+640 & 1) -1800, 2) -1000 \\
MRK279 & 1) -450, 2) -280 \\
RXJ135515+561244 & 1) -990, 2) -780, 3) -220, 4) -110, 5) -0 \\
PG1404+226 & 1) -290, 2) -20, 3) +150 \\
PG1411+442 & 1) -50, 2) +80 \\
NGC5548 & 1) -700, 2) -480 \\
MRK817 & 1) -4100 \\
MRK290 & 1) -220 \\
MRK876 & 1) -3800 \\
MRK509 & 1) -400, 2) -290, 3) -10, 4) +120, 5) +200 \\
IIZW136 & 1) -1500, 2) +20 \\
AKN564 & 1) -250, 2) 0 \\
IRASF22456-5125 & 1) -350, 2) -150, 3) +100, 4) +150, 5) +350 \\
MR 2251-178 & 1) -2000 or -300 \\
NGC7469 & 1) -1900 \\
\enddata
\normalsize
\end{deluxetable}

\end{document}